\documentstyle[psfig,amssym]{elsart}

\begin{document}

\begin{frontmatter}

\title{PKS 2155--304 --- a source of VHE $\gamma$-rays} 

\author{P.M.~Chadwick}, \author{K.~Lyons}, \author{T.J.L.~McComb},
\author{S.McQueen}, \author{K.J.~Orford}, \author{J.L.~Osborne},
\author{S.M.~Rayner}, \author{S.E.~Shaw},
\author{K.E.~Turver\thanksref{corauthor}}, \author{G.J.~Wieczorek}

\address{Department of Physics, Rochester Building, Science Laboratories,
University of Durham, Durham, DH1 3LE, U.K.}

\thanks[corauthor]{Corresponding author}

\begin{abstract}

The close X-ray selected BL Lac PKS 2155--304 has been observed using
the University of Durham Mark 6 very high energy (VHE) gamma ray
telescope during 1996 September/October/November and 1997
October/November. VHE gamma rays with energy $ > 300$ GeV were detected
with a time-averaged integral flux of $(4.2 \pm 0.7_{stat} \pm
2.0_{sys}) \times 10^{-11}~{\rm cm}^{-2}~{\rm s}^{-1}$. There is
evidence for VHE gamma ray emission during our observations in 1996
September and 1997 October/November. The strongest emission was detected
in 1997 November, when the object was producing the largest flux ever
recorded in high-energy X-rays and was detected in $> 100$ MeV
gamma-rays. The VHE and X-ray fluxes are correlated.

\end{abstract}


\end{frontmatter}

\section{Introduction}

X-ray selected BL Lacs (XBLs) at small redshift are sources of very high
energy (VHE) gamma rays at energies above 300 GeV. The BL Lac first
detected as a source of VHE gamma rays was Mrk 421 \cite{punch1992},
which is an EGRET gamma-ray source. Mrk 501 is also a source of VHE
gamma rays \cite{quinn1996}, although not detected at GeV energies with
EGRET, and it exhibits low-level emission with flaring
\cite{catanese1997a}. The BL Lac 1ES 2344+514 emits episodic VHE
gamma rays \cite{catanese1997b,catanese1998}. All of these objects have
small redshift ($z \sim 0.03$). There have been no reported detections
of VHE gamma rays from radio-selected BL Lacs (RBLs --- see e.g.
\cite{kerrick1995,roberts1998}).

Stecker et al. \cite{stecker1996} have interpreted the gamma ray results
in the GeV -- TeV range and propose a model in which RBLs will be GeV
gamma ray sources and XBLs will be TeV sources. On the basis of this
model, PKS 2155--304, despite having a redshift of 0.117, is predicted
to be a strong TeV gamma-ray source. PKS 2155--304 was discovered as an
X-ray source during observations made with the {\it HEAO-1\/} satellite
\cite{schwartz1979,griffiths1979} and PKS 2155--304 may be regarded as
the archetypal X-ray selected BL Lac object. It has a history of rapid,
strong broadband variability. In 1997 November, contemporaneous with
some of the observations reported here, X-ray emission was detected with
the {\it Beppo-SAX\/} satellite \cite{chiappetti1997} with a flux equal
to the strongest previous outburst, and GeV gamma rays were detected
with EGRET \cite{sreekumar1997}.

\section{Observations}

The University of Durham Mark 6 atmospheric \v{C}erenkov telescope has
been in operation at Narrabri, NSW, Australia since 1995
\cite{armstrong1997}. PKS 2155--304 was observed in 1996
September/October/November and 1997 October/November under moonless,
clear skies. Data were taken in 15-minute segments. Off-source
observations were taken by alternately observing regions of sky which
differ by $\pm~15$ minutes in right ascension from the position of PKS
2155--304. We have a total of 41 hours of on-source observations with an
equal quantity of off-source data.

\section{Results}

Events considered suitable for analysis are those which are confined
within the sensitive area of the camera (i.e. within $1.1^{\circ}$ of
the centre of the camera) and which contain sufficient information for
reliable image analysis, i.e. which have {\it SIZE} $ > 500$ digital
counts, where 3 digital counts $\sim$ 1 photoelectron, and 200 digital
counts are produced by a 125 GeV gamma ray. 

\begin{figure}

\centerline{(a)}
\centerline{\psfig{file=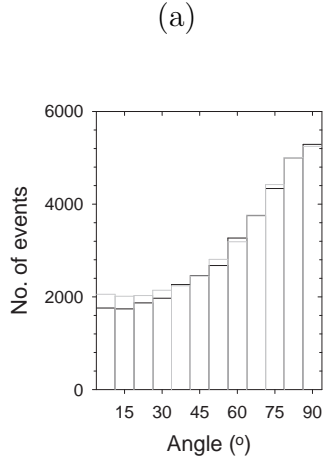,height=6cm}}
\centerline{(b)}
\centerline{\psfig{file=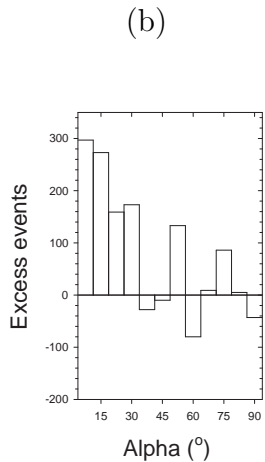,height=6cm}}

\caption{(a) The $ALPHA$ distributions ON and OFF source for
PKS 2155--304. The gray line refers to ON source data. (b) The
difference in the {\it ALPHA} distributions for ON and OFF source
events. \label{alpha_plot} }

\end{figure}

The \v{C}erenkov image can be parameterized using techniques developed
by the Whipple group which describe both the shape and the orientation
of the image. In addition a measure of the fluctuations between the
centroids of the samples recorded by the left and right flux collectors
of the Mark 6 telescope provides a further discriminant
\cite{chadwick1998a}. Gamma rays are identified on the basis of image
shape and left/right fluctuation, and then plotting the number of events
as a function of the pointing parameter {\it ALPHA}; $\gamma$-ray events
from a point source will appear as an excess of events at small values
of {\it ALPHA} ($< 22.5^\circ$). The number of events remaining ON and
OFF source after the application of the standard selections are
summarized in Table \ref{result_table}. The {\it ALPHA} distribution for
the whole dataset is shown in Fig. \ref{alpha_plot}. No normalization of
ON and OFF data rates has been applied. 


\begin{table*}

\caption{The results of various event selections for the PKS 2155--304
data. Data from observations at all zenith angles $< 60^\circ$ have been
combined.}
 
\label{result_table}

\begin{center}

\vspace{0.1cm}
\begin{tabular}{@{}lrrrr}

\hline
& On & Off & Difference & Significance \\
\hline
Number of events & 1021083 & 1023280 & $-2197$ & $-1.5~\sigma$ \\
\\
Number of size and & 600856 & 598733 & 2123 & $1.9~\sigma$ \\
distance selected events & & & & \\
\\
Number of shape & 37125 & 36151 & 974 & $3.6~\sigma$ \\
selected events & & & & \\
\\
Number of shape and & 6099 & 5370 & 729 & $6.8~\sigma$ \\
{\it ALPHA} selected events & & & & \\
\hline
\end{tabular}

\end{center}

\end{table*}


\subsection{Observed flux}

544 excess events identified as $\gamma$-rays were detected in 32.5
hours of on-source observation at zenith angles less than $45^{\circ}$.
The current selection procedure for data recorded at zenith angles $ <
45^{\circ}$ is estimated to retain between 20 and 50\% of the original
gamma ray events. We have assumed a value of 20\%. The collecting area
has been estimated from Monte Carlo simulations and is $\sim 5.5 \times
10^{8}~\rm{cm}^{2}$ for these observations. Using this estimate, the
$\gamma$-ray flux incident on the earth's atmosphere was found to be
$(4.2 \pm 0.75_{stat} \pm 2.0_{sys}) \times 10^{-11}~{\rm cm}^{-2}~{\rm
s}^{-1}$ for a threshold $> 300$ GeV.

\subsection{Time variability}

\begin{figure}

\centerline{\psfig{file=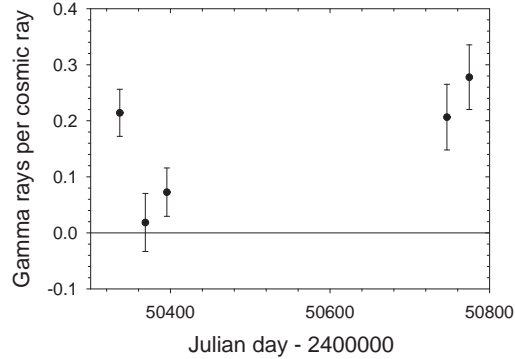,height=6cm}}

\caption{The variation of the VHE gamma ray flux from PKS 2155--304
averaged over the observing periods in 1996 September, October and
November and in 1997 October and November.  \label{monthly} }

\end{figure}

We show in Fig. \ref{monthly} the variation in the detected $\gamma$-ray
flux for observations in 1996 September, October and November and in
1997 October and November. The strength of the $\gamma$-ray emission is
defined as the number of $\gamma$-ray candidates (shape and orientation
selected events) to the number of background protons during the same
observation. This makes allowance for variations in the sensitivity of
our telescope as the zenith angle changes. Using the test for constancy
of emission used by the EGRET group \cite{mclaughlan1996}, the data
suggest that the VHE $\gamma$-ray emission is time variable.


\section{Discussion}

PKS 2155--304 is the fourth X-ray emitting BL Lac to be established as a
VHE emitter and, thus far, the most distant TeV emitter from earth with
a redshift of 0.117. The emission shows the features of time variability
demonstrated by the other X-ray selected BL Lacs detected at VHE
energies. 

\begin{figure}

\centerline{\psfig{file=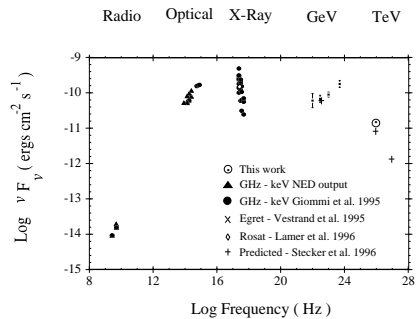,height=6cm,angle=90}}

\caption{The spectral energy distribution of PKS 2155--304. The VHE
point is from the present work. Other data is from the work of
\cite{giommi1995,lamer1996,vestrand1996} and the prediction is from
\cite{stecker1996}. \label{spectrum} }

\end{figure}

In Fig. \ref{spectrum} we show the spectral energy distribution (SED)
from PKS 2155--304 from radio to VHE gamma ray energies, including the
present results. The SED is consistent with those for the other
VHE-emitting blazars. The VHE behaviour is in general agreement with the
predictions of Ghisellini et al. \cite{ghisellini1998} and, as for the
other three established VHE blazars, indicates the success of the
unified model of AGNs in explaining the gross features of VHE emission. 


\begin{figure}

\centerline{\psfig{file=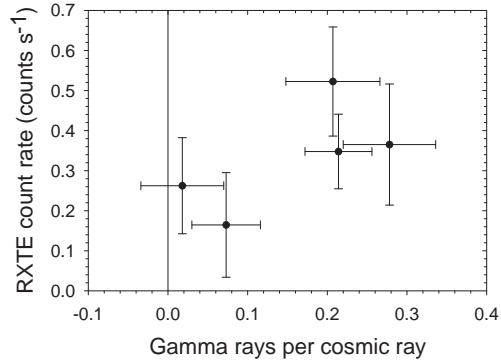,height=6cm}}

\caption{The correlation between VHE emission from PKS 2155--304 and the
X-ray flux measured by the {\it RXTE} satellite when averaged over an
observing period ($\sim 10$ days); the correlation coefficient is
$0.66$ (4 df). \label{xrays} }

\end{figure}

During our observations of PKS 2155--304 measurements of the daily
average of its 2 -- 10 keV X-ray emission were available from the ASM on
{\it RXTE}.\footnote{Available on the web at
\mbox{http://space.mit.edu/XTE/asmlc/pks2155-304.html}.} In Fig.
\ref{xrays} we show the relation between the average VHE emission during
each of the five months and the X-ray rate from the {\it RXTE} quick
look analysis averaged over the days when TeV observations were made.
The correlation coefficient for the data is 0.66 for 4 degrees of
freedom (df). The general correlation between X-ray and VHE emission in
PKS 2155--304 is in agreement with multiwavelength observations of Mrk
421 and Mrk 501 (reviewed by e.g. Ulrich et al. \cite{ulrich1997}).
These suggest that the same population of electrons produces the X-ray
and TeV emission. The strongest VHE emission from PKS 2155--304 occurred
in 1997 November (during a multiwavelength campaign). We note that the
strongest X-ray emission ever observed from this object was detected by
{\it Beppo-SAX} during observations in 1997 November 22 -- 24
\cite{chiappetti1997}. EGRET also detected GeV gamma-rays, again at a
flux considerably higher than previous detections, during an observation
from 1997 November 11 -- 17, just before the onset of our November
observation \cite{sreekumar1997}. A recent analysis of {\it RXTE} and
EGRET data \cite{vestrand1998} suggests that our 1997 November VHE gamma
ray observations were preceded by a strong flare.


\ack

We are grateful to the UK Particle Physics and Astronomy Research
Council for support of the project. This paper uses quick look results
provided by the ASM/{\it RXTE} team and uses the NASA/IPAC Extragalactic
Database (NED) which is operated by the Jet Propulsion Laboratory,
Caltech, under contract with the National Aeronautics and Space
Administration.

\end{document}